# UNDERSTANDING MODERN INTRUSION DETECTION SYSTEMS: A SURVEY

[1]LIU HUA YEO, [2]XIANGDONG CHE, [3]SHALINI LAKKARAJU

[1,2,3]College of Technology, Eastern Michigan University, United States
Email: [1]lyeo@emich.edu, [2]xche@emich.edu, [3]slakkara@emich.edu

**Abstract:** Intrusion detection systems (IDS) help detect unauthorized activities or intrusions that may compromise the confidentiality, integrity or availability of a resource. This paper presents a general overview of IDSs, the way they are classified, and the different algorithms used to detect anomalous activities. It attempts to compare the various methods of intrusion techniques. It also describes the various approaches and the importance of IDSs in information security.

## 1. INTRODUCTION

Intrusion detection is defined as identifying unauthorized use, misuse and abuse of computer systems by both inside and outside intruders. The main task of an intrusion detection system (IDS) is to defend a computer system or computer network by detecting hostile attacks on a network system or host device (Khan Pathan, 2014), monitoring the events occurring in a computer system or network and analyzing them for signs of intrusions. Security incidents resulting from attempted attacks violate the CIA triad of computer security; Confidentiality, Integrity and Availability. The National Institute of Standards and Technology (NIST) defines intrusion as an attempt to compromise CIA or to bypass the security mechanisms of a computer network (Bace and Mell, 2001).

An IDS identifies, logs and reports possibly security incidents. It is the software or hardware system to automate the intrusion detection process and is typically placed inline, at a spanning port of a switch, or on a hub in place of a switch (Pappas, 2008).

## 2. IDEAL INTRUSION DETECTION SYSTEM

### 2.1. Materials and Procedures

An ideal intrusion system should address the issues below, regardless of the mechanism it is based on, to defend against attacks and intrusions. Some of them are described below (Vinchurkar & Reshamwala, 2012):

- The system should be able to run continuously without human supervision. It must be reliable enough to be run in the background of the system being observed.
- It should not be a "black box", which means its internal workings should be examinable from outside.
- It must be fault tolerant, which means that it must survive a system crash and not have its database rebuilt at restart.
- It must resist destruction. The system can monitor itself to ensure that it has not been subverted.
- It should observe and record deviations from normal behavior.
- It must be easily tailored to the system. Every system has a different usage pattern, and the defense mechanism should adapt themselves easily to these patterns.
- It must deal with changing system behavior over time as new applications are being added.
- The system must have a very low false negative and false positive rate.

Since a typical IDS generates a large amount of traffic and events in its logs, the key is for the system to only generate alerts on events of interest. An effective IDS has a low rate of false positives and false negatives.

Table 1 - True positives and negatives

|  | POSITIVE | NEGATIVE |
|---|---|---|
| **TRUE** | Alerts when there is malicious traffic | Silent when traffic is benign |
| **FALSE** | Alerts when traffic is benign | Silent when malicious traffic occurs |

The ideal tuning of an IDS, therefore, maximizes the instances of true positives and true negatives.

## 3. CLASSIFICATION OF INRUSION DETECTION SYSTEMS

There are many different ways to classify the various types of IDS in a production network. These classifications are not mutually exclusive; for instance, a network-based IDS may be using the signature-based approach to detection. The following diagram describes the most common methodologies of IDS classifications, although the list is certainly not exhaustive.

### 3.1. Host-based IDS

A host-based IDS (HIDS) is an IDS that generally operates within a computer, node or device. Its main function is internal monitoring, although many variants of HIDS have been developed that can be used to monitor networks (Singh & Singh, 2014). Primarily, it monitors and analyzes the internals of a computer, node or device. An HIDS determines if a system has been compromised and warns administrators accordingly (de Boer & Pels, 2005). For example, it can detect a rogue program that accesses a system's resources in a suspicious manner, or discover that a program has modified the registry in a harmful way.

HIDSs were the first types of intrusion detection software to have been designed (Gupta, 2015). Unlike network-based IDSs, an HIDS can inspect the full communications stream. NIDS evasion techniques, such as fragmentation attacks or session splicing, do not apply because the HIDS is able to inspect the fully recombined session as it is presented to the operating system (Hay & Cid, 2008). Encrypted communications can be monitored because an HIDS inspection can look at the traffic before it is encrypted. This means that HIDS signatures will still be able to match against common attacks and not be blinded by encryption.

An HIDS is also capable of performing additional system level checks that only IDS software installed on a host machine can do, such as file integrity checking, registry monitoring, log analysis, rootkit detection, and active response (Hay & Cid, 2008).

### 3.2. Network-based IDS

A network-based IDS (NIDS) differs from an HIDS in that it is usually placed along a LAN wire. It attempts to discover unauthorized and malicious access to a LAN by analyzing traffic that traverses the wire to multiple hosts. There are many algorithms for detecting malicious traffic, but they generally read inbound and outgoing packets and searches for any suspicious patterns. Any alert generated by an NIDS allows it to notify administrators or take active actions such as blocking the source IP address.

Three of the most common placements of NIDS are directly connecting it to a switch spanning port, using a network tap, and connected inline.

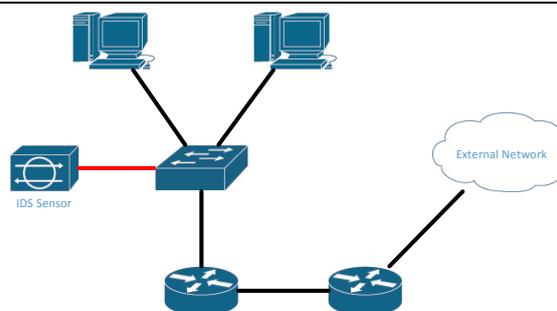

Figure 1 - IDS using a switch spanning port

Figure 3 shows the IDS connected to a switch that has SPAN port configuration capability. On some managed switches, a SPAN port can be configured to send all packets on the network to that port as well as their ultimate destination (Baker, Beale, Caswell & Poor, 2004). In this configuration, the switch copies all traffic it receives to the IDS interface being used to monitor traffic. The major downside of this method is increased bandwidth and resource usage, since the switch must work twice as hard to deliver traffic.

Very few modern LANs use hubs due to the lack of security. Hubs allow systems to intercept traffic not intentionally sent to them. When using either a hub or switch with SPAN port capabilities, the systems on the internal network are not at the mercy of the IDS having a system failure brining the network down (Pappas, 2008). Making use of a switch SPAN port is a common method of connecting sensors.

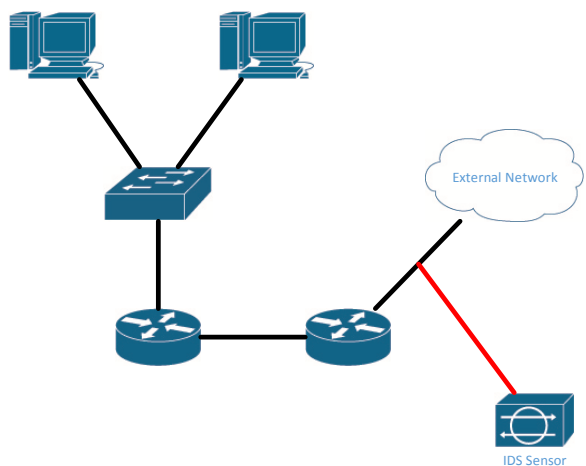

Figure 2 - IDS using network tap

Figure 4 shows an IDS using a network tap, which essentially replicates data passed through the wire. Network taps are not commonly found in typical computer networks but may be purchased (Pappas, 2008). Taps are handy when a network administrator needs to setup a hasty monitoring solution, perhaps to troubleshoot a problem or temporarily deploy an IDS (Pappas, 2008). Overall, a network tap is needed when the network does not have managed switches, is not using hubs, or when putting an IDS inline is impractical.

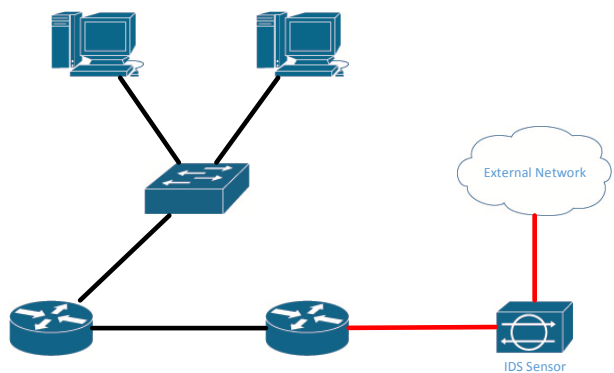

Figure 3 - IDS connected inline

Figure 5 illustrates an IDS connected inline. This instance includes two connections, shown in red, with one connected to the uplink port of the switch, and the second connected to the external network. In most cases, this is not the best method to use because system failure of the IDS will prevent systems on the internal network from communicating with external systems (Pappas, 2008).

However, the benefit of the inline configuration is a guarantee all packets will be seen by the IDS. Packets are subject to being missed when an IDS is connected to a switch SPAN port, especially when that switch is busy processing a large burst of traffic. Missed traffic may be lost forever if they were not captured by a network sniffing protocol. Depending on the capability of an inline IDS, a similar burst may lead to congestion of network performance.

Although an NIDS is a powerful monitoring system for network traffic, there are several disadvantages (Hay & Cid, 2008). Common NIDS evasion techniques such as fragmentation attacks, session splicing, and even denial-of-service (DoS) attacks can be used to bypass an NIDS, rendering it useless. If the communications between hosts are encrypted, a passive NIDS does not have the ability to unencrypt a message in transition.

### 3.3. Signature-based Detection

In the signature-based approach, an IDS looks for packets and compares them with predefined rules or patterns known as signatures that are defined in the database. These attack signatures pass specific traffic or activity that is based on known intrusive activity (Gupta, 2015). The main advantage of this technique is the simple and efficient processing of audit data. Signature-based approaches have a much lower rate of false positives. On the other hand, the very nature of signature-based detection means that such an approach is ineffective against zero-day attacks for which there may not be a discovered ruleset or established method of attack yet. With the rate of new attacks and malicious activities every hour, a signature-based IDS is only as good as the recency of its signature database and rulesets.

### 3.4. Anomaly-based Detection

Anomaly-based IDS works by identifying patterns from users or groups of users already defined. This approach looks for variations and deviations from an established baseline behavior which might indicate malice. It involves increased amount of processing which is used by anomaly detector for studying the behavior of the system from its audits (Cannady & Harrell, 1996). The baseline must first be created of the system, network or program activity. This baseline is the profile of what a normal scenario, usage, bandwidth or behavior would look like in a specific network environment. Thereafter, any activity that deviates from the baseline is treated as a possible intrusion (Gangwar & Sahu, 2014) and an alert would be generated.

The biggest advantage of anomaly-based approach is its ability to detect zero-day attacks, since it does not depend on an established signature database, but merely deviations from an established baseline. The behavior of each target system or network is unique, therefore anomaly-based approaches use customized profiles which in turn make it difficult for an attacker to know with certainty what activity it can carry off without setting off an alarm.

On the other hand, anomaly-based IDSs have a high false positive rate. It also requires time to establish a baseline behavior when it is first placed in a new network environment or host device. These systems are also more complex and the difficulty of associating an alarm with the specific event that triggered that alarm (Gupta, 2015).

Table 2 - Comparison between different IDS classifications

| Type | Advantages | Disadvantages |
|---|---|---|
| HIDS | 1. More accurate in intrusion detection 2. Able to detect encrypted attacks. 3. Does not require additional hardware | 1. Higher cost. 2. May cause performance issues or resource hogging. |
| NIDS | 1. Low cost. 2. Detect network-based attacks such as | 1. High fluctuations in network traffic cause packets to be lost. |

|  | | | | |
|---|---|---|---|---|
|  |  | denial-of-service attacks. | 2. | Requires more CPU power and resources in a large-scale LAN. |
|  |  |  | 3. | Unable to analyze encrypted packets. |
| **Anomaly-based** | 1. | Ability to detect zero-day attack attempts. | 1. | Slow to work when placed in a new environment. |
|  | 2. | Low false negative rate. | 2. | High false positive rate. |
|  |  |  | 3. | Low detection rate for known attacks. |
| **Signature-based** | 1. | High response time for known attacks. | 1. | Limited capability to detect zero-day attacks. |
|  | 2. | Low false positive rate. | 2. | Signature database must be updated frequently. |

### 3.5. Passive and Active IDS

When classifying IDSs, we can also categorize them by the way IDSs respond during an attack. A passive IDS merely records, analyzes, logs and alerts an administrator about the possibility of an attack. In terms of placement, passive IDSs are generally placed off to the side in a network.

An active IDS can take actions when it detects a possible intrusion, such as blocking further traffic from a specific network source or locking down the system with safe mode. In modern security systems, an active IDS is also known as an Intrusion Prevention System (IPS). IPSs are placed inline in a network.

#### 3.5.1 Intrusion Prevention Systems (IPS)

The main function of an IPS is to intervene in cases of suspected attacks. Generally, an IPS is essentially a combination of access control devices – such as firewalls and routers – and IDSs. In other words, an IPS is an IDS with access control capabilities or active response methods.

Like IDSs, an IPS can be host-based or network-based, and uses anomaly-based detection (prevention) or a signature- or ruleset-based approach.

The following are common countermeasures implemented by IPSs:
- Denying the traffic. This is the simplest method, where the intrusion system blocks the IP addresses and ports involved – both source and destination. The downside to this method is that many devices on the global network are hidden behind a global address. Blocking that address will also block other legitimate traffic that may be located behind that address.
- Active logging. Although logging is a feature shared by IDSs, an IPS can increase the usability of a log by, for example, automatically exporting traffic logs that meet certain criteria to external network analysis software such as Wireshark.
- Communicating with a separate device with access control capabilities. Many modern IDSs and IPSs also complement the operations of a LAN by communicating with an external, or separate, firewall or router, which have access control capabilities. In the event of an intrusion, an IDS/IPS can send an alert or request to a firewall or router. The firewall or router will then take the necessary actions to deal with the intrusion, such as dropping the packets or blocking further traffic from that source.
- Sending a TCP reset (Carter, 2005). If an attack is a TCP-based attack, an IPS can send a reset signal back to the attacker's protocol stack, which would close the current session, and can be repeated as frequently as needed.
- Setting an SNMP trap (Burns, Adesina & Barker, 2012). When an alarm is triggered, the intrusion system will send an SNMP trap to indicate to an SNMP management system that a network or device is under attack. The management system can choose to take an action based on the event, such as polling the agent directly, or polling other associated device agent to get a better understanding of the event.

#### 3.5.2 IPS or IDS?

Although the market trend is focusing on IPS rather than IDS with the advancement of DDoS attacks (Fuchsberger, 2005), there are reasons to choose between an IPS and an IDS. IDS rarely cause latency in network traffic, because it is generally off to the side (unless placed inline) and all traffic are merely copied to the sensor. An IPS, on the other hand, may cause small delays in traffic because its inline placement means that every packet has to be inspected and analyzed before being forwarded to its destination.

If an intrusion system is disabled, for example by accident or power outage, an IDS will not cause denial of service due to its positioning. An IPS, which would have to be connected inline, would negate the availability of network resources. Modern

IPSs have preventive mechanisms such as fault tolerance or backup power to minimize disruption to network activities.

If cost is an issue, many routers such as Cisco® routers allow specific modules or firmware to be installed on top of the existing routers to provide intrusion detection and prevention capabilities.

## 4. DETECTION TECHNIQUES

From different sources, systems like rule-based expert systems, state transition analysis, and genetic algorithms are direct and efficient ways to implement signature detection. Inductive sequential patterns, artificial neural networks, statistical analysis and data mining methods have been used in anomaly detection. There are different kinds of frameworks used for anomaly-based detection.

This section presents an extensive study over the various intrusion detection classifier techniques and hybrid detection techniques. A few proposed methods could be described as follows.

### 4.1. Bayesian Networks

Bayesian networks are probabilistic graphical models that represent sets of variables and their probabilistic independencies. Bayesian theory is named after Thomas Bayes. His theory can be explained as follows:

If the events $A_1$, $A_2$, ... and $A_n$ constitute a partition of the sample space S such that $P(A_k) \neq 0$ for $k = 1, 2, ..., n$, then for any event B such that $P(B) \neq 0$:

$$P(A_i|B) = \frac{P(A_i \cap B)}{P(B)} = \frac{P(A_i)P(B|A_i)}{\sum_{k=1}^{n} P(A_k)P(B|A_k)} = \frac{P(A_i)P(B|A_i)}{P(B)}$$

Darwiche, et. al (2010) stipulated that Bayesian networks have been used in many computer science fields, such as email spam filters, speech recognition and pattern recognition, because of their ability to build coherent results by using probabilistic information about a specific situation.

Bayesian networks are directed acyclic graphs where the nodes represent variables and whose edges encode conditional dependencies between those variables (D. Heckerman, 1995). These are applied to anomaly detection in so many ways; for example, Valdes et al. has developed an anomaly detection system that employed naive Bayes, which is a two-layer Bayesian network that assumes complete independency between the nodes.

### 4.2. Genetic Algorithm (GA)

GA is a search technique that is used to find an appropriate solution to search problems. Genetic algorithms have been applied in anomaly detection in many ways, as they are flexible and a powerful search method. Some network intrusion detection approaches have used genetic algorithms for the classification of instances, while others like fuzzy data mining approach have applied this technique for feature selection. To list out an advantage of GA, it selects the best feature and has better efficiency but its method is complex.

### 4.3. Inductive Rule Generation Algorithms

These algorithms are one of the most famous techniques used. In this technique, we have a predictive model decision tree that maps observations of an item to conclusions about the item's target value.

The decision tree (DT) is very powerful and popular data mining algorithm for decision-making and classification problems. It is also used in many real-life applications like medical diagnosis, radar signal classification, weather prediction, credit approval, and fraud detection. This decision tree can be constructed from large volume of dataset with many attributes, because the tree size is independent of the dataset size. It can process both numerical and categorical data but trees created from numeric datasets can be complex. Construction of inductive rule generation algorithms may not require any domain knowledge. It can handle high dimensional data and the representation is easy to understand (R. Patel, A. Thakkar and A. Ganatra, 2012). However, it is limited to one output attribute. Decision tree algorithms are unstable and most decision tree construction methods are non-backtracking,

### 4.4. Outlier Detection

Outlier detection approach is based on the idea of semi-supervised learning in which the system would learn a baseline data, and consider any instances that do not fit in the normal data profile as an anomaly.

Most of the anomaly detection algorithms require a set of baseline data to train the model, and they assume that anomalies can be treated as patterns never observed before. Since an outlier is defined as a data point which is very different from the rest of the data, so based on some measure, we employ several outlier detection schemes to see how efficiently these schemes may deal with the problems of anomaly detection. In statistics-based outlier detection techniques, the data points are modeled using a stochastic distribution and these points are determined to be outliers depending upon their relationship with this model.

### 4.5. Clustering

This technique is based on two important assumptions (L. Portnoy, E. Eskin, & S.J. Stolfo., 2001). First, majority of the network connections represent normal traffic and only a very small percentage of that traffic is malicious. And second, malicious traffic is statistically different from normal traffic. Anomalies will be detected based on their cluster size, i.e., large clusters are meant to be baseline data, and the rest correspond to malicious attacks.

Clustering is unsupervised learning. Labeling the data is not necessary and natural patterns in the data are extracted. It does not require the use of a labeled data set for training.

### 4.6. Neural Networks

Neural networks are networks of computational units that jointly implement complex mapping functions. First, the networks are trained with a labeled data set. Testing instances are then fed into the network to be classified as either normal or anomalous. An example of the neural network technique which is widely used in anomaly detection is the Support Vector Machines (SVM) (S. Mukkamala, G. Janoski, and A. Sung, 2002).

This method would be effective if the exact characteristics of the attack are already known. However, these intrusions are constantly changing because of the individual approaches taken by the attackers and regular changes done in the software and hardware of the targeted systems. Because of the wide variety of attacks and attackers despite their dedicated effort to constantly update the rule base of an expert system can never hope to accurately identify the variety of intrusions.

For these constantly changing natures of these network attacks, we require a flexible defensive system that can analyze these enormous amounts of network traffic in a manner, which is less structured than rule-based systems. For example, a neural network-based signature detection system could potentially address many of the problems that are found in rule-based systems. The inherent speed of neural networks is another benefit of this approach, as it requires a timely identification of attacks, and the processing speed of the neural networks could enable intrusion responses to be conducted before irreversible damage could be done to the system. It has a high signal-to-noise ratio and requires more time and more sample-training phase.

### 4.7. Fuzzy Logic

Fuzzy logic starts and builds on a set of user-supplied human language rules. The fuzzy systems convert these rules into their mathematical equivalents. This simplifies the job of the system designer and the computer, and results in a much more accurate representation in the way systems behave in the real world. Fuzzy logic is also simple and flexible. It can handle problems arising from imprecise and incomplete data, and model nonlinear functions of arbitrary complexities. Fuzzy logic techniques have been employed in the computer security field since the early 90's (Hosmer, 1993). Its ability to model complex systems makes it a valid alternative in the computer security field to analyze continuous sources of data and even unknown or imprecise processes (Hosmer, 1993).

Fuzzy logic has the potential in the intrusion detection field when compared to those systems using strict signature based matching or classic pattern deviation detection. Bridges and Vaughn (2000) state that the concept of security itself is fuzzy. And in other words, the concept of fuzziness helps to smooth out the abrupt separation of normal behavior from abnormal behavior. This means a given data point falling outside a defined baseline interval, will be considered anomalous to the same degree regardless of its distance from the interval. Fuzzy logic has a capability to represent imprecise forms of reasoning in areas where firm decisions must be made in indefinite environments like intrusion detection.

Dokas et al. (2002) suggested a model that works by building rare class prediction models for identifying known intrusions and their variations and anomaly/outlier detection schemes for detecting novel attacks whose nature is unknown.

Researchers propose techniques to generate fuzzy classifiers using genetic algorithms that can detect anomalies and some specific intrusions. The main idea was to evolve two rules; one for the normal class and other for the abnormal class using a baseline profile data set.

### 5. COMPARISON OF TECHNIQUES

In this section, we look at how the various intrusion detection techniques compare, and in what situations are they best suited for. Each technique or approach to programming the system to recognize intrusion or outright threats have different variables involved for defining the exclusionary or inclusionary criteria for defining what constitutes a threat. Our review of each technique poses a challenge in determining suitable techniques based on the system's specific needs and awareness of threats in relationship to risk assessment. In the business world, how the technique will be applied to the system directly relates to the purpose and function of the system to meet business needs. While this is simplistic, there is also thematic attachment to deriving application of the technique within specific and definable parameters of the system.

However, and because of global business implication to managing the firm's technology as an asset and understanding how the firm needs interactions with unknown clients, there is a need for basing technique upon flexible patterns of information sharing. Biermann, Cloete, and Venter (2001) regard a means of devising clear intent in terms of each techniques drawbacks and advantages of use. The system must define the technique and compare them based upon availability, utility, integrity, authenticity, confidentially, and possession (Biermann, Cloete, & Venter, 2001).

Hubballi and Suryanarayanan (2014) find that part of the issue with defining the parameters of the IDS and therefore strategically determining the specific technique and purpose, also rides upon how well each technique can recognize events without error. They contend that a signature-based design may be ineffective in the constant need to rewrite parameters to re-identify the threat, such that "discovery of new flaws and vulnerabilities occur continuously, to write good signatures one needs to have complete understanding of the behavior and also sufficient data to analyze. Due to this dependency, this method is always error prone."

Such a technique may not allow for gray areas and therefore should be replaced by fuzzy logic where parameters are protected by filtering for what may be depicted as an error. The signature may not fit with the rule based logic applied or within the signature, or the rules are too stringent and therefore, information packets are seen as threats when instead they are not (false positivity). Rewriting the rule causes a continuous action and need for the system to learn from itself rather than specific directions/rules set in place.

Another relationship between techniques to consider is how similar these structures function within the design of the system to point to specific correlative formats for alarms where they are seen in normative and rules-based functionality. Bayesian algorithms capture the same essence as the signature approach where rules-based definitions of threats also apply. Xiao et al. (2014, p. 128) comment that the reason for choosing the Bayesian derived techniques coincides with robust modeling and "joint distribution of random variables and reasoning under uncertainty."

Even with this logic, Xiao et al. (2014) find weakness in the node function of the technique and how these nodes present opportunities for false alarms just as found in the signature based technique. Continued application of knowledge and loading the patterns to seek in defining the level of safe with uncertain requires continued revision of the system. Xiao et al. (2014) determine that utilizing Bayesian algorithms at the core of the IDS also means adopting further tools within the design toward a hybrid model. Hybrid augmentation seeks a dual platform for filtering the information based upon extensions where dynamic coding for applying human behavior modelling opens up functionality for greater leverage between identifying factors of certainty and uncertainty. Such further design based upon the Bayesian approach continues the robustness of the framework but allows for analytics to be applied in a way that considers profoundly shifting the variables for non-threat and threat definers. Taking a step further to analyze the definition allows for greater functionality and less false alarms.

Differences in approach bordering on hybrid design also focuses upon fuzzy logic and clustering techniques. However, clustering and fuzzy logic can only be used in situations where there is a broad basis for data analysis and there is the expected element of learning from the data. These forms of IDS techniques would not work well in situations where the needs of the administrator are simply to block rules based packets of information. Systems defined by a set of prerequisites as a single layer of function would not need the depth or level of precision implied from the use of the clustering or fuzzy logic. Indeed, such systems would not need the hybrid approach either but may function best with Bayesian or signature based techniques as the points of identifiers are clear.

Fuzzy logic and clustering allow for further dissemination and critical review of data as the application is in real time which works better for soft computing or cloud systems (Ashfaq et al., 2017). Lin et al. (2015) determine that the use of clustering proves desirable because it can integrate similar themes or traits with the timing of receiving the data with the needs of the system and its activities. Integration serves to provide the system with clear messages of what information is partitioned, where, and why. Due to the level of detail, situations that call for a simpler solution will not justify investing in such intricate design features unless the activity level or purpose of the administrator's actions warrant such protections (Mitrokotsa & Dimitrakakis, 2013).

Kabir, Onik, and Samad (2017) remark that the weakness of the Bayesian approach is due to its popularity and widespread usage as the preferred method. This points to the need for a flexible hybrid or dual platform where learning contributes to the evolution of the system. Therefore, Kabir et al. (2017) find that Bayesian programming is a strong foundation but that ultimately technique design should move towards featuring clusters for anomalous behaviors, where one part determines the "true" user while the other determines normal patterns of behavior of that user.

With the migration toward soft computing and cloud data storage systems, IDS and techniques need to be focused on providing classifiers for cross validation of data sources where there is little time discrepancy. The technique must have parameters for all time temporalities (Mitrokotsa & Dimitrakakis, 2013). The concept of robustness remains central to the design of a technique that meets the needs of business functionality and multi-dimensions of time. This calls for a technique based in robustness, sturdy and strict in rules-based procedures but also learning from uncertain information, such as possible false positives.

Wang et al. (2010) support the use of fuzzy clustering to maneuver uncertainty for classifying attacks to the system as the business needs grows more toward application of risky interactions. Being fuzzy means the data is uncertain or falling into an undefinable realm per the rules based system and even anomaly definition. The gray area of data will find similar traits, common threads, and reasoning for being grouped together. The fuzzy cluster hybrid as a technique finds strength in weaving between levels and sharing, and learning from the information of the system rather than falsely believing that the information is a threat. Issues with this approach pertain to installing hybrid forms of security measures where the data must unlock rules to move within the system. Wang et al. (2010) demonstrate how this may create limitation where there are walls to block information exchange but without further clustering to the result, the threat will be blocked or partitioned off to serve the needs of one group defined by the cluster. Wang et al. (2010) find how this can provide further design features and usefulness for the user in terms of who has access to the data and who does not. If data is found to be a threat, then because of the clustering effect, the system is protected but the data is still analyzed on some level of value to the firm. What proves to be beneficial is the ability to learn from the data while keeping it separated from other sensitive and proprietary data as intellectual property (Ashfaq et al., 2017).

## 6. CURRENT TRENDS

As the world moves towards cloud computing, mobile applications and wireless networks, information security is more crucial than ever. A survey report by the Computer Security Institute found that 45.6 per cent of respondents reported that they had been subject to at least one targeted attack in the past year (Richardson, 2011). Data breaches has cost companies billions of dollars in damages and lost assets. The biggest threats to network security are disgruntled former employees and attacks from outside the company (Schneider, 2012).

The field of intrusion detection and prevention is still in its infancy (Hunt & Zeadally, 2012). By using probing tools, some attackers explore a victim's network prior to launching an attack. A sophisticated IDS may be able to correlate data obtained from the attacker's reconnaissance – possibly along with additional log data – to either forecast the attack or to obtain better forensic evidence during or after the attack (Hunt & Zeadally, 2012). However, although some progress has been made recently with distributed IDS architectures, many IDSs cannot detect complex intrusions and distributed or coordinated attacks (Zhou, Leckie & Karunaseckera, 2010).

## 7. CONCLUSION

We have introduced an overview of the different types of intrusion detection systems, approaches, methodologies and techniques for IDSs. Each technique and class of IDS has its superiority and limitations, so we should be mindful when selecting the best approach. We compared and contrasted each technique and approach to determine which works best in a particular situation. We focused our study on the most common intrusion detection models such as NIDS and HIDS, and both the anomaly- and signature-based approach to detection. Finally we also presented current trends in the world of information and network security in the last several years and the direction that information technology is heading to.

★ ★ ★